\documentclass[12pt]{iopart}

\usepackage{xcolor}
\begin{document}

\title[On the breakdown of space-time in general relativity]{On the breakdown of space-time in general relativity}

\author{Farrukh A. Chishtie$^{1,2,3}$}
\address{$^{1}$Department of Mathematics, The University of Western Ontario, London, ON N6A 3K7, Canada}
\address{$^{2}$Department of Occupational Science and Occupational Therapy, University of British Columbia, Vancouver, BC V6T 2B5, Canada}
\address{$^{3}$Peaceful Society, Science and Innovation Foundation, Vancouver, BC V6S 1K3, Canada}
\ead{fachisht@uwo.ca}
\vspace{10pt}
\begin{indented}
\item[]March 2023
\end{indented}

\begin{abstract}
Based on the canonical quantization of $d>2$ dimensional general relativity (GR) via the Dirac constraint formalism (also termed as `constraint quantization'), we propose the loss of covariance as a fundamental property of the theory. This breakdown occurs for the first-order Einstein Hilbert action, whereby loss of diffeomorphism invariance, besides first class constraints, second class constriants also exist leading to non-standard ghost fields which render the path integral non-covariant. We also attempt, for the first time, the canonical quantization via calculation of the path integral for the equivalent Hamiltonian formulation of GR for which only first class constraints exist. However, loss of covariance still occurs in this action due to loss of diffeomorphism invariance and structures arising from non-covariant constraints in the path integral. In contrast, we find that covariance as a symmetry is restored and quantization with perturbative calculations is possible in the weak limit of the gravitational field of these actions. Hence, we firstly establish, for the first time, that the breakdown in space-time is a property of GR itself as its limitation, indicating it to be an Effective Field Theory (EFT). We further propose that the breakdown of space-time occurs as a non-perturbative feature of GR in the strong field limit of the theory. Besides GR, we also note that covariance is preserved when constraint quantization is conducted for non-Abelian gauge theories, such as the Yang-Mills theory. These findings are novel from a canonical gravity formalism and EFT approach, and are consistent with GR singularity theorems, yet are general and extend them, as the singularity theorems indicate breakdown at a strong field limit of GR in black holes. Our findings are in contrast to the asymptotic safety program. They support emergent theories of space-time and gravity, though are unique, as they do not require thermodynamics such as the entropic gravity program. From an EFT view, these indicate that new degrees of freedom and/or principles in the non-perturbative sector of the full theory are a requirement, whereby covariance as a symmetry is broken in the high energy (strong field) sector of GR. 
\end{abstract}



\maketitle

\section{Introduction}
Quantum Gravity (QG) as represented by a theory which explains and predicts gravitational phenomena across quantum mechanical scales is being developed in a diversity of ways including via quantum field theoretical approaches. One approach is the quantization of general relativity (GR) theory which expected to lead to a full and coherent theory of gravitation at quantum mechanical scales, and this is an ongoing and active area of research. Dirac laid the foundation by quantizing GR via a novel constraint formalism [1,2]. Pirani et al. [3] formulated a Hamiltonian formulation of the GR action based on this approach, and Dirac later developed another Hamiltonian-based GR (HGR) action aimed to quantize gravity via quantum field theoretical approaches [4]. Einstein [5] also formulated GR into what is now known as the first order (1EH) Einstein-Hilbert action. 

In this work, and on the basis of recent attempts to quantize GR via the Dirac constraint formalism and path integral approach (which we will term as `constraint quantization'), we present that loss of covariance in quantization is present in both the full 1EH and HGR formulations, though different canonical constraint structures arise in their respective actions. We establish, for the first time, that this breakdown in space-time indicates that GR is an Effective Field Teory (EFT), as via the same constraint quantization successful quantization is possible in the weak limit of the gravitational field, which indicates that covariance is only upheld in this limit. Hence, the breakdown in space-time is a non-perturbative property of GR (for $d>2$) itself, indicating its limitation at higher energy quantum mechanical scales. 

In the following section, we present a summary of results derived in [8-10],[12,13] and next, a first attempt of canonical quantization of the HGR action in [3,4]. We find that for the HGR action, covariance is broken, as while there are not only explicit non-covariant first class constraints, but there is breakdown of covariance in the diffeomorphism constraints as well, as covariance is upheld for these constraints only the surface of the primary constraints. We further present a counterexample regarding the successful quantization of non-Abelian gauge Yang Mills theories, for which first class constraints exist, and for which covariance is maintained and is not dependent only on the surface of primary constraints [17]. Based on these findings, we propose that the loss of manifest covariance in GR using the constraint quantization approach is due to its application to the full action. We demonstrate and establish, that GR is actually an EFT, while this breakdown is a non-perturbative property of $d>2$ dimensional GR theory itself from the recovery of covariance from linearized versions (which have different constraint structures for 1EH and HGR actions [20-22]). Our findings therefore, are in contrast with the asymptotic safety program which assumes GR to hold across all energy scales [19]. We also connect our findings to recent work focusing on 1EH and second order (2EH) actions using background and Lagrange multiplier fields and the path integral approach [23-30]. We further note that our findings are consistent with existing work on space-time breakdown in GR in the case of black holes by Penrose [36,37], whereby this breakdown in GR happens in the strong limit of the gravitational field of a black hole, and also with Effective Field Theory (EFT) results which hold at one loop order in the low-energy limit of the theory [29, 30]. We also find that space-time is an emergent feature, however there is loss of covariance in the high-energy (strong field limit) of the theory, and that this property does not require thermodynamics as is required by entropic gravity [31]. We finally provide three novel findings based on the results in this work, as a way forward for the development of a comprehensive theory of QG.        

\section{Canonical quantization of $d>2$ dimensional GR via canonical constraint formalism}

The $d$-dimensional 1EH action is defined as:

\begin{eqnarray}
\mathcal{L}_d^{1\,EH}=h^{\mu\nu}\left(G^{\lambda}_{\mu\nu,\lambda}+\frac{1}{d-1}G^{\lambda}_{\lambda\mu}G^{\sigma}_{\sigma\nu}-G^{\lambda}_{\sigma\nu}G^{\sigma}_{\lambda\nu}\right)
\end{eqnarray}   

where $h^{\mu\nu}=\sqrt{-g}g^{\mu\nu}$, $g_{\mu\nu}$ is the metric, $G^{\lambda}_{\mu\nu,\lambda}=\Gamma^{\lambda}_{\mu\nu}-\frac{1}{2}(\delta^{\lambda}_{\mu}\Gamma^{\sigma}_{\sigma\nu}+\delta^{\lambda}_{\nu}\Gamma^{\sigma}_{\sigma\mu})$ and $\Gamma^{\lambda}_{\mu\nu}$ is the affine connection. This choice of this action is based on the property that it is trinomial in power of the metric, thereby rendering perturbation theory possible via the Feynman diagrammatic approach. In contrast, in the second order (2EH) action, increasing powers of the metric tensor in perturbation thereby rendering calculation via Feynman diagrams and renormalization impossible [6, 7]. In [8-10], for $d>2$, using the Dirac constraint formalism, it was found that besides first class constraints, second class constraints also existed. For constrained systems with second class constraints, the generating functional via the path integral formalism derived by Senjanovic [11] is:
\begin{eqnarray}
Z[J] = \int D\Phi\, D\Pi\, D\lambda_a\, D\kappa_a \det \left\lbrace \phi_a, \chi_b\right\rbrace \mathrm{\det}^{1/2}\left\lbrace \theta_a, \theta_b\right\rbrace  \delta(\chi_b)\nonumber\\
\times \exp i \int dx \bigg(\Pi \frac{\partial}{\partial t}\Phi - \mathcal{H}_c (\Phi, \Pi) - \lambda_a\phi_a(\Phi ,\Pi) - \kappa_a \theta_a (\Phi, \Pi) +  J\Phi\bigg)
\end{eqnarray}
Here $\Phi$ are the canonical fields, $\Pi$ are the conjugate momenta, $ \mathcal{H}_c$ is the canonical Hamiltonian, $\chi_b$ is the gauge condition associated with the first-class constraints $\phi_a$, $\lambda_a$ and $\kappa_a$ are lagrange multipliers, while $\theta_a$ and $\theta_b$ are second-class constraints. Chishtie and McKeon [10] found that these second class constraints were non-covariant for quantization of the theory, while the quantity $p^iq_i-H_c$ and diffeomorphism constraints, which are associated with the gauge symmetry of the 1EH GR action, were also found to be non-covariant. Hence, this loss of manifest covariance in the path integral (Eq. 2) rendered quantization impossible. The non-standard ghost fields were also found in the case of non-Abelian Yang Mills scalar tensor theory by Chishtie and McKeon [12]. 

In this work, we attempt the canonical quantization for the action of the Hamiltionan-based GR (HGR) [3,4]. The HGR action is as follows:
\begin{eqnarray}
\mathcal{L}_d^{H}=\sqrt{-g}g^{\alpha\beta}\left(\Gamma^{\mu}_{\alpha\nu}\Gamma^{\nu}_{\beta\mu}-\Gamma^{\nu}_{\alpha\beta}\Gamma^{\mu}_{\nu\mu}\right)\nonumber \\
\,\,\,\,\,\,\,\,\,\,=\sqrt{-g}B^{\alpha\beta\gamma\mu\nu\rho}g_{\alpha\beta,\gamma}g_{\mu\nu,\rho}
\end{eqnarray}   

where $B^{\alpha\beta\gamma\mu\nu\rho}=g^{\alpha\beta}g^{\gamma\rho}g^{\mu\nu}-g^{\alpha\mu}g^{\beta\nu}g^{\gamma\rho}+2g^{\alpha\rho}g^{\beta\nu}g^{\gamma\mu}-2g^{\alpha\beta}g^{\gamma\mu}g^{\nu\rho}$. 

For this action, it was found that the constraints are first class by Kiriushcheva et al. [13] for $d>2$ dimensions. For these types of constraints, the following path integral by Fadeev [14] applies:

\begin{eqnarray}
Z[J] = \int D\Phi\, D\Pi\, D\lambda_a \,\det \left\lbrace \phi_a, \chi_b\right\rbrace \delta(\chi_b)\times \\ \nonumber
\exp i \int dx \left(\Phi \frac{\partial\Pi}{\partial t} - \mathcal{H}_c (\Phi, \Pi) - \lambda_a\phi_a(\Phi ,\Pi) + J\Phi\right)
\end{eqnarray}   

The $d$-dimensional first class constraints were derived fully in [13] and which are both primary and secondary constraints. With $\pi^{\mu\nu}$ denoting the momenta conjugate to $g^{\mu\nu}$, the primary constraints are: 

\begin{eqnarray}
\phi^{0\sigma}=\pi^{0\sigma}-\frac{1}{2}\sqrt{-g}B^{\left(  \left(
0\sigma\right)  0\mid\mu\nu k\right)  }g_{\mu\nu,k}, \label{eqn4}%
\end{eqnarray}

while the secondary constraints are:
\\
\begin{eqnarray}
\kappa^{0\sigma}  &  =-\frac{1}{2}\frac{1}{\sqrt{-g}}\frac{g^{0\sigma}}{g^{00}%
}I_{mnpq}\pi^{mn}\pi^{pq}\label{eqn8}\\
&  +\frac{1}{2}\frac{g^{0\sigma}}{g^{00}}I_{mnpq}\pi^{mn}A^{\left(  pq0\mid
\mu\nu k\right)  }g_{\mu\nu,k}+\delta_{m}^{\sigma}\left[  \pi_{,k}%
^{mk}+\left(  \pi^{pk}e^{qm}-\frac{1}{2}\pi^{pq}e^{km}\right)  g_{pq,k}\right]
\nonumber\\
&  -\frac{1}{8}\sqrt{-g}\left(  \frac{g^{0\sigma}}{g^{00}}I_{mnpq}B^{\left(
\left(  mn\right)  0\mid\mu\nu k\right)  }B^{\left(  pq0\mid\alpha\beta
t\right)  }-g^{0\sigma}B^{\mu\nu k\alpha\beta t}\right)  g_{\mu\nu,k}%
g_{\alpha\beta,t}\nonumber\\
&  +\frac{1}{4}\sqrt{-g}\frac{1}{g^{00}}I_{mnpq}B^{\left(  \left(  mn\right)
0\mid\mu\nu k\right)  }g_{\mu\nu,k}g_{\alpha\beta,t}\left[  g^{\sigma
t}\left(  g^{00}g^{p\alpha}g^{q\beta}+g^{pq}g^{0\alpha}g^{0\beta}-2g^{\alpha
q}g^{0p}g^{0\beta}\right)  \right. \nonumber\\
&  -g^{\sigma p}\left(  2g^{00}g^{q\alpha}g^{t\beta}-g^{00}g^{\alpha\beta
}g^{qt}+g^{\alpha\beta}g^{0q}g^{0t}-2g^{q\alpha}g^{0\beta}g^{0t}-2g^{t\alpha
}g^{0\beta}g^{0q}+2g^{qt}g^{0\alpha}g^{0\beta}\right) \nonumber\\
&  \left.  +g^{0\sigma}\left(  2g^{\beta t}g^{\alpha p}g^{0q}-2g^{p\alpha
}g^{q\beta}g^{0t}-2g^{pq}g^{t\beta}g^{0\alpha}+2g^{pt}g^{q\beta}g^{0\alpha
}+g^{pq}g^{\alpha\beta}g^{0t}-g^{tp}g^{\alpha\beta}g^{0q}\right)  \right]
\nonumber\\
&  -\frac{1}{4}\sqrt{-g}g_{\mu\nu,k}g_{\alpha\beta,t}\left[  g^{\sigma
t}\left(  g^{\alpha\mu}g^{\beta\nu}g^{0k}+g^{\mu\nu}g^{\alpha t}g^{0\beta
}-2g^{\mu\alpha}g^{k\nu}g^{0\beta}\right)  \right. \nonumber\\
&  +g^{0\sigma}\left(  2g^{\alpha t}g^{\beta\mu}g^{\nu k}-3g^{t\mu}g^{\nu
k}g^{\alpha\beta}-2g^{\mu\alpha}g^{\nu\beta}g^{kt}+g^{\mu\nu}g^{kt}%
g^{\alpha\beta}+2g^{\mu t}g^{\nu\beta}g^{k\alpha}\right) \nonumber\\
&  \left.  +g^{\sigma\mu}\left(  \left(  g^{\alpha\beta}g^{\nu t}%
-2g^{\nu\alpha}g^{t\beta}\right)  g^{0k}+2\left(  g^{\beta\nu}g^{kt}-g^{\beta
k}g^{t\nu}\right)  g^{0\alpha}+\left(  2g^{k\beta}g^{\alpha t}-g^{\alpha\beta
}g^{kt}\right)  g^{0\nu}\right)  \right] \nonumber\\
&  -\frac{1}{2}\sqrt{-g}g^{00}E^{pqt\sigma}\left(  \frac{1}{g^{00}}%
I_{mnpq}B^{\left(  \left(  mn\right)  0\mid\mu\nu k\right)  }g_{\mu\nu
,k}\right)  _{,t}-\frac{1}{2}\sqrt{-g}B^{\left(  \left(  \sigma0\right)
k\mid\alpha\beta t\right)  }g_{\alpha\beta,kt}\nonumber
\end{eqnarray}
\\

where $A^{\alpha\beta0\mu\nu k}=B^{\left(  \alpha\beta0\mid\mu\nu k\right)  }%
-g^{0k}E^{\alpha\beta\mu\nu}+2g^{0\mu}E^{\alpha\beta k\nu}$,
$E^{\mu\nu\gamma\sigma}=e^{\mu\nu}e^{\gamma\sigma}-e^{\mu\gamma}e^{\nu\sigma
}$,
$e^{\mu\nu}=g^{\mu\nu}-\frac{g^{0\mu}g^{0\nu}}{g^{00}}$
and 
$I_{mnpq}=\frac{1}{d-2}g_{mn}g_{pq}-g_{mp}g_{nq}$.\\

Based on these first class constraints, the Castellani procedure [15] was applied in [13] to derive generators $G$ which provide the gauge transformation for the HGR action. While the authors demonstrate covariance of the overall gauge transformation, however, this holds only on the surface of the primary constraint, that is when the condition $\phi^{0\sigma} = 0$ is applied. In particular, we want to specify that the field transformations for the space-space components of the field $g^{\mu\nu}$ which were derived by dropping the second term containing the primary constraint, in the following expression in [13]:   

\begin{eqnarray}
\delta g_{nm}=\{G,g_{nm}\}=\frac{\delta}{\delta\pi^{nm}}\left(
\xi_{\sigma}\kappa^{0\sigma}-\xi_{\sigma}\frac{2}{\sqrt{-g}}I_{abpk}\pi
^{ab}\frac{g^{\sigma p}}{g^{00}}\phi^{0k}\right).
\end{eqnarray}

For the purposes of quantization of the HGR action, the primary constraints must not be restricted in this manner, as this condition applies only for classical GR. Without the assumption of application of the constraint condition, the second term in Eq. 7 leads to non-covariance of the gauge transformation. Hence, for the quantization of the HGR action, the gauge contributions $\chi_b$ are non-covariant in Eq. 4, which leads to non-quantization of the HGR action. Since this leads to the same finding as for 1EH action, we deduce that loss of covariance occurs in GR theory via the constraint quantization procedure.  

As demonstrated above, while there is loss of covariance in the full GR actions with the application of the Dirac constraint formalism, however this symmetry is restored in the weak limit of these actions. Linearized GR actions are derived by expanding the field metric tensor around the flat metric as follows,  

\begin{eqnarray}
h^{\mu\nu}=\eta^{\mu\nu}+\kappa\phi^{\mu\nu},
\end{eqnarray} 

where $h^{\mu\nu}=h^{\nu\mu}$, $\eta^{\mu\nu}$ is the flat metric ($\eta^{\mu\nu}=diag(+++...-)$ and $\phi^{\mu\nu}$ is the perturbation around this metric.  

In [16-18], the constraints derived for linear (weak-field) 1EH action and HGR action are shown to be different from those derived from their respective full theory actions, though the path integral is covariant as a result. This restoration of covariance in the weak field limit of GR is a key part of our proof, which relies on the EFT framework besides conventional QFT. Here, an EFT framework provides us with the distinction of the weak field limit (or linear gravity) associated with a lower energy scale as compared to the strong field limit which is associated at the higher energy scale. With this finding, we therefore explicitly show for the first time that GR is actually an EFT (using canonical quantization results for GR in both its weak and strong limits). Using the EFT framework complements conventional QFT, whereby the distinction between weak and strong limits of fields as degrees of freedom are made via their evolution with respect to energy scales as elements that we draw upon as part of our proof. Hence, the recovery of covariance in the weak limit of the metric tensor is an important finding, whereby covariance symmetry is restored and field path integral quantization is indeed then made possible. This finding is another indication that the loss of manifest covariance in the full GR actions is not an artifact due to the Dirac constraint quantization procedure. Moreover, this result also highlights how our finding is different from and contrasts completely from, for example, the asymptotic safety scenario/program as a theory of QG by Weinberg et al. [19], assumes GR to be valid across all energy scales. In this scenario, asymptotic safety therefore implies, \textit{a priori} that the gravitational coupling has an ultraviolet (UV) fixed point or the theory is UV safe (which means the $\beta$-function of the coupling vanishes in the high energy limit). Our findings, in contrast, indicate that the gravitational coupling encounters a singularity in the strong field limit (high energy limit) at a certain energy scale, indicating that GR is actually an EFT, with the loss of manifest covariance in the high energy limit as an indication of the limitations of this theory.  

Interestingly, the canonical structures for the $d=2$ for 1EH and HGR actions are different from the $d>2$ dimensions as found in [20-22]. As shown in these studies, the $d=2$ case displays unique properties including the presence of an addtional $SO(2,1)$ symmetry in the Hamiltonian. This leads to a considerably different behaviour from $d>2$ GR as the constraints (including symmetries) themselves render the theory without any independent degrees of freedom. Given the focus on dynamical GR in our work, we aim to address issues of quantization of $d=2$ case in a future work [33]. However, there is no loss of generalization of our present findings in not considering this case, as this is a special case which has no bearing on physical implications or applications of GR, given its non-dynamical nature.    

Recently, and in further support of our findings for restoration of covariance in the weak limit of GR theory, McKeon and collaborators et al. [23-30] have used background field theory and perturbation approach in the weak limit of the gravitational field to render the 1EH and 2EH actions renormalizable, finite and restricted them to one-loop order using background and lagrange multiplier fields. These are also consistent with EFT results whereby quantum corrections are possible at 1-loop order in the low-energy limit of 4-dimensional GR, however there is no indication of breakdown in covariance in this approach, except that renormalizability beyond one-loop in this formalism is not possible [31,32]. In this regard, [30] aims to derive a one-loop Beta function based on the quantized and renormalized 1EH action, which would indicate the extent to which GR holds in the weak limit and strong dynamics begin. We intend to follow this work with this investigation of the dynamics responsible for this onset of breakdown in covariance [33] and also coupling the 1EH action with Yang-Mills theory using the canonical formalism [34]. 

Besides our main proof, we finally note that no known quantized gauge theories (e.g. Yang-Mills theory) have the property of gauge transformations derived on the surface of primary constraints, thereby preserving covariance. Hence, the non-covariance of the diffeomorphism constraints and gauge transformations explicitly demonstrate that quantization of the full HGR action is not possible. To illustrate this, while both of the full EH actions indicate breakdown in covariance, in [17], as a counter example, it was shown that application of the Dirac constraint and the path integral formalism in fact does recover covariance for gauge theories such as Yang Mills theory, though the space and time components are separated at the start of the analysis. So, for Yang-Mills theory for a vector field $A^{a}_{\mu}$
\begin{eqnarray}
\mathcal{L}^{YM}=-\frac{1}{4}F^{a}_{\mu\nu}F^{a\mu\nu}
\end{eqnarray}  
with the covariant derivative, $D^{ab}_\mu=\partial_{\mu}\delta^{ab}+f^{apb}A^{p}_{\mu}$ and $[D_{\mu},D_{\nu}]^{ab}=f^{apb}F^{p}_{\mu\nu}$, it is found that the constraints are first class. Upon application of the path integral for first class systems, the following covariant path integral is obtained:
\begin{equation}
Z[J]=\int DA^{a}_{\mu}\det(-\partial^{\mu}D^{ab}_\mu)\exp i \int dx\left[-\frac{1}{4}F^{a}_{\mu\nu}F^{a\mu\nu}-\frac{1}{2\alpha}(\partial\cdot A^{a})^2+J^{a\mu}A^{a}_{\mu}\right]
\end{equation} 

With this counter example, we want to further emphasize that the constraint quantization does indeed work for covariant gauge theories such as Yang-Mills theory. This example is another clear indication the breakdown of covariance in the case of $d>2$ dimensional GR is actually a property of the theory itself, rather than an artifact of the Dirac constraint formalism. The provision of this counter example is an illustration of the usefulness and success of the Dirac quantization procedure on gauge theories which differ from GR. Overall, our results and findings indicate that GR is not totally intractable at quantum mechanical scales; rather, in the weak field limit of the metric tensor, it is shown to yield different canonical constraint structure than the full theory, whereby covariance is restored rendering constraint quantization possible in this limit. 

\section{Findings and Implications}  

Our findings of the loss of covariance or breakdown in space-time as a non-perturbative feature of the GR theory are consistent with the singularity theorems by Penrose which demonstrate the breakdown of the theory in the strong limit of the field in black holes [35,36]. Our present findings supports QG scenarios whereby space-time and gravity is emergent without requiring thermodynamics, as is required by entropic gravity [37]. This theory is also known as emergent gravity, whereby it is assumed here that gravity is actually an entropic force caused by quantum-level disorder, which shows homogeneity at the macro-scale. Therefore, gravity is not a fundamental interaction in entropic gravity. There are fundamental issues with entropic gravity as it has been shown to violate the First Law of Thermodynamics on ordinary space-time surfaces [38]. In contrast, our findings which have no thermodynamic assumptions circumvents these issues, with GR as an EFT, and therefore we offer an alternative scenario of emergent gravity. Based on the findings from constraint quantization, we therefore arrive at three novel findings postulated as follows:

1. The breakdown of space-time in General Relativity is a non-perturbative property indicating that it is an EFT, whereby covariance is lost in the high energy limit (strong field limit) of the theory. \\
2. Covariance as a symmetry is restored in the low energy (weak limit) of the GR theory, indicating it as non-fundamental and space-time (and gravity) as emergent properties.\\
3. Physical principles more fundamental than covariance and/or new degrees of freedom and dynamics are required for a coherent and comprehensive theory of Quantum Gravity.

\section{Conclusions}
In this work, we propose three novel findings based on the canonical structures and the loss of manifest covariance when quantization is conducted for various formulations of $(d>2)$-dimensional GR. We firstly propose that loss of manifest covariance property of the theory itself (valid for $(d>2)$ dimensions) rather than an artifact of the Dirac constraint procedure and our proof demonstrates, for the first time, that GR is actually an EFT. These are deduced from different canonical constraint structures for the GR 1EH action and the Hamiltonian formulation of GR, whereby there is loss of covariance in both actions of GR happens, when constraint quantization is attempted. Our proof is completed with the finding that covariance is restored in the weak limit of the actions (though the canonical structure is different from the full GR actions) whereby quantization is achieved in this limit of the theory. This indicates space-time (and gravity) as an emergent property, without requiring thermodynamic conditions as is required in entropic gravity. Finally, based on these novel findings, we further propose that in the onset of non-perturbative dynamics in the high-energy (strong field limit), new degrees of freedom and/or physical principles are required for the development of a full theory of QG as space-time (covariance) via GR are no longer valid across all energy scales of this theory.

\section{Acknowledgments}
I thank Gerry McKeon for the helpful discussions on this work and fruitful collaborations over the years which inspired this work. I would also like to thank the anonymous reviewers for their helpful questions and useful recommendations.

\section{Competing interests}
 The author declares that there are no competing interests.

\section{Data Availability Statement}
Due to its mathematical and theoretical physics basis, no data is available to share for this work.

\end{document}